\documentclass{PoS}

 \title{Why things fall}

\ShortTitle{Why things fall}

\author{\speaker{Olaf Dreyer}\\%
        Center for Theoretical Physics, Massachusetts Institute of Technology,\\
        77 Massachusetts Ave, Cambridge, MA 02139   \\
        E-mail: \email{odreyer@mit.edu}}

\abstract{ In this paper we discuss Internal Relativity, a recent program to address the problem of quantum gravity. In our approach we change the relationship between spacetime and matter. Currently we view matter as propagating on spacetime. Einstein's equations encode how spacetime curves due to the presence of matter and how spacetime, in turn, tells matter how to propagate. In internal realtivity matter and spacetime cease to exist as distinct entities, rather, they arise simultaneously from an underlying quantum system. It is through the emergent matter degrees of freedom that geometry is inferred.  We have termed our program Internal Relativity to stress the importance of looking at the system from the point of view of an internal observer. We argue that special relativity is then a natural consequence of this viewpoint. The most important new aspect of Internal Relativity involves how gravity appears. It is not just a new quantum theory of gravity but a new theory of gravity. We also argue that the presence of a massive object implies curvature. In particular we show that Newtonian gravity arises in the appropriate limit. Our argument implies that there is no propagation without gravitation. }

\FullConference{From Quantum to Emergent Gravity: Theory and Phenomenology\\
                 June 11-15 2007\\
                 Trieste, Italy}

\begin{document}

\section{Introduction}\label{sec:intro}
One of the most interesting aspects about the search for a quantum theory of gravity is that the goal of the search is not clear, leading to a number of different approaches to the problem. One can distinguish these different approaches by the r\^ole gravity plays in the fundamental theory. In research programs such as loop quantum gravity \cite{tt}, spin foams \cite{sf}, and causal dynamical triangulations \cite{loll}, the aim is to quantize the gravitational field. Gravity is an integral part of the formulation of the theory. In other approaches to quantum gravity the gravitational field is not part of the initial formulation of the theory but instead emerges. An example is the low energy behavior of Fermi liquids as investigated by Volovik \cite{volovik}. In the special case in which Ffermi points are present he shows that the system can be described by an emergent dynamic metric. This metric degree of freedom is truly emergent in that the fundamental theory consists only of interacting fermions and has no gravitational degrees of freedom. String theory can also be viewed as falling into this second category. In its initial formulation,  gravity appears as a spin-2 vibrational mode of the string, while in more recent formulations, such as the AdS/CFT correspondence, gravity is conjectured to appear in the appropriate limits.  

Here we shall follow an approach of the later kind, i.e., where gravity is not part of the initial formulation of the theory but is emergent. We shall differ from the other approaches in this category in the way the gravitational field emerges. In string theory and in Fermi liquids, gravity appears as a distinct low energy excitation, in this case a massless spin-2 excitation, however, additional matter degrees of freedom also appear.  In our approach,  there is no such clean distinction between matter and gravity degrees of freedom.  Instead we are taking seriously the fact that we only know geometry through matter. Only by using matter, the proverbial clocks and rods, can we infer geometry; geometry alone is not accessible to us. In Internal Relativity we only look at the available matter degrees of freedom and ask what geometry we obtain if we only use these degrees of freedom. It is important here that we do not include information, like an absolute time, that is only available to an observer external to the system. To stress this point we have termed our program \emph{Internal Relativity} \cite{dreyerinoriti}.

What then is the geometry that we will find? We will show in this paper that generically the geometry is that of a curved manifold with a Lorentzian signature. That one finds a Lorentzian signature is not so surprising, or new. It was by asking similar questions to the ones we are asking here that Lorentz found his transformations, in a step that was the beginning of special relativity. The central new contribution from Internal Relativity involves the emergence of gravity. We show that actual objects that could function as rods and clocks automatically have a gravitational mass. We establish this by showing that Newtonian gravity appears in the quasi-static limit. 

The logic of Internal Relativity is thus as follows. We start with a quantum mechanical system. It has no gravitational degrees of freedom and it is not obtained from the quantization of a classical theory. We then find the ground state of the system together with its low lying excitations.  We use  these excitations to construct rods and clocks. We then argue that these clocks and rods will feel the force of gravity by showing that Newton's law of gravity applies to them. The resulting geometry is thus a Lorentzian curved geometry. It is currently a conjecture of Internal Relativity that if one continues in this direction, adhering to a strictly internal point of view, one will find the Einstein equations. 

The organization of this article is as follows. In section \ref{sec:ir}, we review the basics of Internal Relativity from \cite{dreyerinoriti}. We stress that it is important to ask how a system looks to an observer inside the system. This point of view is essential for the emergence of relativity. In the following section \ref{sec:examples}, we give a number of examples for Internal Relativity. We start with the early history of special relativity and discuss how Lorentz arrived at the transformations that now carry his name. Lorentz employed his arguments in a classical setting. Using simple examples from solid state physics we show that they can just as easily be applied in a quantum setting. In section \ref{sec:gravity}, we present our argument why gravity is automatically present in the kind of systems we look at. The main result of this section is the derivation of Newton's law. In the process we derive a new expression for the gravitational mass of a bound object. Since the class of theories we are looking at are quantum mechanical,   the emerging theory is automatically a quantum theory of gravity. This section establishes that the geometry encountered by the internal observers is not just flat Minkowski space but instead is a curved Lorentzian manifold. To complete the Internal Relativity program it remains to be shown that the equivalence principle is in fact true. We discuss in section 4 how far along we are in this. The premise of Internal Relativity is very different than the usual one and it brings a new perspective and possible new answers to long standing problems in physics, such as the cosmological constant problem and the problem of time, as we discuss in the final section \ref{sec:disc}. 

\section{Internal Relativity}\label{sec:ir}
In the introduction we stated that we intend to derive general relativity without  putting it into the fundamental theory or finding it in the form of an emergent massless spin-2 excitation. We plan to start with a theory that possesses a preferred time, is not relativistic or background independent, and find general relativity. How can this be done? The key principle that will allow us to do this is to not look at the system from the outside but instead to ask what the system looks to an observer from the inside. It was exactly this point of view that marked the beginning of special relativity. Given that the world is governed by the Maxwell equations, the question that Lorentz asked was what this meant for {\em our} ability to measure space and time intervals. As we shall see in more detail in the next section, what he found was length contraction and time dilation. That is, starting from a theory formulated in Newtonian absolute space, Lorentz found Minkowski space by using the internal point of view.

It is the basic idea of Internal Relativity to return to this more physical attitude towards relativity. The difference is that this time around we are looking for more than special relativity. We claim that the internal point of view has not been taken far enough. If one strictly adheres to it, one finds not only special relativity but also general relativity. This is the central novelty of Internal Relativity. Furthermore, we use quantum mechanical systems from the start. The theory that we will ultimately arrive at is thus a quantum theory of gravity. 

The reason that we expect this to be true is the dual r\^ole of matter. Not only do we find geometric notions like space and time intervals through matter, but, also, through propagation and interactions, matter has inertial properties. In \cite{dreyerinoriti}, we conjectured that it is this dual role that is at the heart of the equivalence principle and Einstein's equations:

\begin{description}
\item[Conjecture] When notions of distance, time, mass, energy, and momentum are defined in a completely internal way the equivalence principle and Einstein's equations hold automatically. 
\end{description}

This is the main claim of Internal Relativity. Because it is so important to what follows, we summarize how it differs from the way we currently understand general relativity: Currently, we think of our world as being composed of two parts, matter and geometry. Matter moves \emph{on} geometry and geometry curves due to the presence of matter on it. This interaction between matter and geometry is described by the Einstein equations. In our view, matter and geometry have a more dual r\^ole. One can not have one without the other. Both emerge from the fundamental theory simultaneously. In the next section we will look at a couple of examples of how special relativity arises in non-relativistic systems. In section \ref{sec:gravity}, we are then ready to see why gravity is also naturally present.

\section{Examples}\label{sec:examples}
After this general introduction into Internal Relativity we shall now look at some specific examples. In this section we focus on the appearance of special relativity and leave gravity for the next section. The first example is a classical one and goes back to the birth of special relativity: it is how Lorentz found the transformations that now carry his name. We shall concentrate on Bell's \cite{bellsr} version of Lorentz's arguments (see also \cite{brown} for a more careful historical analysis). 

\begin{figure}
\begin{center}
\includegraphics[height=4cm]{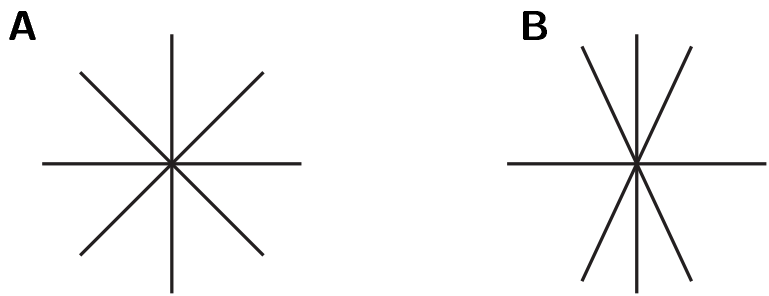}
\end{center}
\caption{ \textbf{\textsf{A}} shows the field of a charged particle at rest. The field is completely spherically symmetric. When the particle is moving the field is no longer spherically symmetric, rather, it is squeezed in the direction in which the particle moves, as shown in \textbf{\textsf{B}}. }
\label{fig:charge}
\end{figure}

Bell starts by considering the field of a charged particle with charge $Z e$ that is moving in the $z$-direction with velocity $v$. It is given by
\begin{eqnarray}
E_x & = & Z e x (x^2 + y^2 + {z^\prime}^2)^{-3/2}\left(1 - \frac{v^2}{c^2}\right)^{-1/2},\\
E_y & = & Z e y (x^2 + y^2 + {z^\prime}^2)^{-3/2}\left(1 - \frac{v^2}{c^2}\right)^{-1/2},\\
E_z & = & Z e z^\prime (x^2 + y^2 + {z^\prime}^2)^{-3/2},\\
 & & \nonumber \\
B_x & = &  -\frac{v}{c} E_y,\\
B_y & = & \frac{v}{c} E_x,
\end{eqnarray}
where 
\begin{equation}
z^\prime = (z - z_N(t))\left(1 - \frac{v^22}{c^2}\right)^{-1/2},
\end{equation}
and $z_N(t)$ is the position of the charge (see figure \ref{fig:charge}). Nowadays we obtain this result by simply taking the field of a point charge and applying a Lorentz boost to it. Before the apparatus of special relativity was available this solution had to be derived laboriously from the field equations; as it was first done by Heavyside.  

From these expressions for the electromagnetic field, Bell argues for Lorentz contraction and time dilation by looking first at a single atom. An electron moving in this field will follow an orbit that is squeezed in the $z$-direction. Furthermore, the period will be lengthened. Both squeezing of the orbit and lengthening of the period involve the factor
\begin{equation}
\gamma = \left(1 - \frac{v^2}{c^2}\right)^{-1/2},
\end{equation}
that makes it appearance in the formulae above. A piece of matter made up of atoms whose electrons show this kind of change in their orbit will then squeeze by the same amount as the orbit of the individual atom. In particular, the measurement devices we use to measure space and time intervals show this same change. To an observer using these devices the world looks like it is governed by special relativity. 

This example illustrates important aspects of Internal Relativity. The fundamental setup of the theory is not relativistic at all. The whole theory is formulated in absolute Newtonian space. Special relativity comes from the \emph{dynamics} of the matter fields. It is the change of the form of the electric field that gives rise to the change in the behavior of measuring devices. Internal observers that only have access to these devices have no way of knowing that, on a fundamental level, there is an absolute space. 

This argument is completely classical and thus of rather limited interest. Our next example is a quantum mechanical one.
To illustrate that the above argument also works in a quntum mechanical setting, we first look at a very simple model: the XY-model in a one-dimensional spin chain. The Hamiltonian of the system is given by:
\begin{equation}\label{eqn:original}
H = \sum_{i=1}^N (\sigma^+_i\sigma^-_{i+1} + \sigma^-_i\sigma^+_{i+1}).
\end{equation}
Here $\sigma^\pm = \sigma^x \pm i \sigma^y$, and the $\sigma$'s are the Pauli matrices. Again, the model is not relativistic. It is discrete and has a preferred time. However, relativity becomes immediately apparent when one looks at the low energy effective Hamiltonian for this model. The above Hamiltonian can be written in terms of a two-component spinor fields $\psi_\alpha$, $\alpha = 1,2$:
\begin{equation}
H = \int dx\ \psi^\dagger(x)\beta i \partial_x\psi(x),
\end{equation}
where
\begin{equation}
\beta = \sigma^1 = \left(\begin{array}{cc}0 & 1\\ 1 & 0\end{array}\right)
\end{equation}
(for details about how this low energy description can be obtained see \cite[chapter 4]{fradkin}). We thus see that the low energy world looks like two-dimensional Minkowski space to internal observers.

This example is, of course, too simple for our purposes. It has only one type of excitations and these particles are  free. More interesting examples with more realistic particle content do exist. A particularly interesting example that is also built using spins on a lattice was proposed by Levin and Wen \cite{wenlevin}. They argue that their model gives rise to QED-like physics at low energies. The details of how special relativity emerges at low energies are discussed in \cite{dm}.

\section{Gravity}\label{sec:gravity}
Having discussed examples of the emergence of special relativity from a non-relativistic quantum systems, we now want to look at the emergence of gravity. The conjecture of section \ref{sec:ir} says that one can not have an inertial mass without at the same time also having a gravitational mass. In this section we see how far we are towards proving this conjecture. The following is a sketch of the argument leading to an expression for the gravitational mass. 

\begin{table}
\begin{center}
\begin{tabular}{ccc}
level 0 & ground state & $\theta_0$\\
 & & \\
 & & \\
level 1 & excitations & $f_k$ \\
 & "elementary particles" & \\
 & & \\
 & & \\
level 2 & bound states & bound objects\\
\end{tabular}
\end{center}
\caption{Our setup is based on these three levels. Level zero is the ground state. It is described by the parameter $\theta_0$. The next level is given by excitations. These are local deviations from $\theta_0$. Bound states of excitations make up level two. }
\label{tab:setup}
\end{table}

We first need  to distinguish three levels of emergent dynamics in our theory. We will call the ground state of our model the {\em  zeroth level}. In the quantum mechanical example from the previous section, the ground state is characterized by a certain value of the order parameter. Let us denote the value of this order parameter by $ \theta_0$.

The next level of emergent dynamics is given by the local deviations from the ground state $\theta_0$, i.e., by the excitations. In our model these are the traveling spin waves or, equivalently, the fermions $f_k$. We call this {\em level one}.  These excitations can be thought of as representing elementary particles in a simplified model of our world.  The fact that they are emergent is not visible to us.

{\em Level two} is  given by bound states of these excitations. Examples of these in our world are solid objects. The important point about level two objects is that they do not leave $\theta_0$ unchanged. Because they are bound objects of excitations and excitations are local deviations from $\theta_0$ the order parameter near a level two object will deviate from $\theta_0$. This fact will become important shortly. Table \ref{tab:setup} gives an overview of the three levels that we have introduced.

\begin{figure}
\begin{center}
\includegraphics[height=5cm]{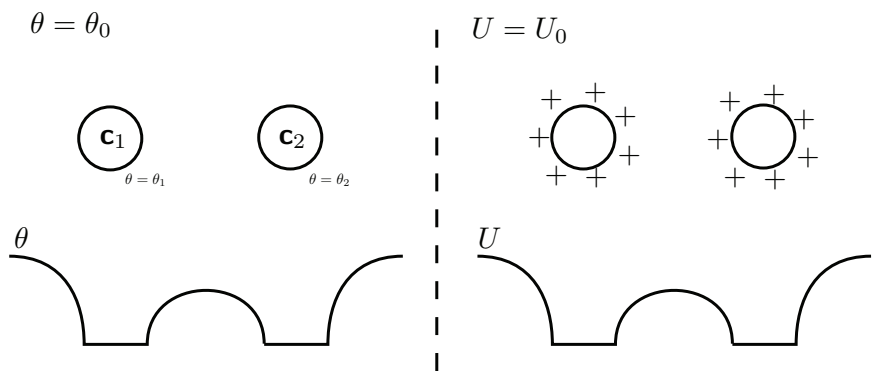}
\end{center}
\caption{Newton's law for classical objects. We argue for Newton's law of gravitation by looking at an analogous situation in electrostatics. The Laplace equation describes both the behavior of the order parameter $\theta$ and the electric potential $U$. The free energy $F$ in the case of the electric potential $U$ is an integral over the square of the electric field $E=\nabla U$, i.e., it is the energy in the electric field.}\label{fig:newton}
\end{figure}

Given this setup, we now want to argue for the natural presence of gravity. We will start by arguing for Newtonian gravity in quasi-static situations, i.e., in the limit of low velocities. Let us look at two bound objects. As we noted in the last paragraph the order parameter $\theta$ will not be uniform; rather it will reflect the presence of the bound objects \textbf{\textsf c}$_1$ and \textbf{\textsf c}$_2$ (see figure \ref{fig:newton}). To calculate the spatial distribution of $\theta$ we look at the free energy $F$ of the system. It will generically include a term of the form
\begin{equation}\label{eqn:freeenergy}
\int d^3x\; \; (\nabla\theta)^2.
\end{equation}
It is this term that is responsible for the propagation of the excitations. By varying this term with respect to $\theta$ we find the equations governing $\theta$:
\begin{equation}\label{eqn:laplace}
\frac{\delta F}{\delta \theta} = 0 \ \ \Rightarrow \ \  \Delta\theta = 0.
\end{equation}
The order parameter $\theta$ thus satisfies a Laplace equation. We can now obtain Newton's law of gravitation in the approximation of low velocities. In this approximation we calculate $F$ for different static situations and then deduce the force on the classical objects by noticing how $F$ varies. To calculate $F$ we solve equation (\ref{eqn:laplace}) for $\theta$ in the presence of the classical objects \textbf{\textsf c}$_1$ and \textbf{\textsf c}$_2$. These impose boundary conditions on $\theta$:
\begin{equation}\label{eqn:boundary}
\theta\vert_{\partial \mbox{\scriptsize \textbf{\textsf c}}_i} = \theta_i,\ i=1,2.
\end{equation}
The boundary condition at infinity is $\theta = \theta_0$, the vacuum value. We can solve these equations explicitly. It is easier though to just compare them to the corresponding equations in electrostatics. If we replace $\theta$ by the electric potential $U$ then equations (\ref{eqn:laplace}) and (\ref{eqn:boundary}) become the equations for a static electric field in the presence of two charged bodies (see again figure \ref{fig:newton}). Expression (\ref{eqn:freeenergy}) for the free energy in turn becomes the energy stored in the electric field. In the electrostatic example the force between the charged bodies will be the Coulomb force, i.e., a force inversely proportional to the square of the distance. By analogy we find the Newton's law of gravity for our situation, i.e., there will be a force of the form
\begin{equation}
F \simeq \frac{m_1 m_2}{r^2},
\end{equation}
where 
\begin{equation}\label{eqn:mass}
m_i \simeq \int_{\partial  \mbox{\scriptsize \textbf{\textsf c}}_i} (\nabla\theta)\cdot d\sigma, \ \ i = 1,2.
\end{equation}

The argument that we have given above applies to all bound objects. What is more we can understand why gravity is only attractive. It is for the same reason that two people in a hammock always end up stuck together in the middle of the hammock or that two grains of dust in the surface of water will attract each other: Since all bound objects alter the free energy in the same way the free energy is minimized when they are close together. This is because when the two objects are far apart the order parameter tries to assume the vacuum value $\theta_0$ between the objects. The required changes in $\theta$ can be omitted when the objects are close to each other, leading to a smaller contribution to the free energy. 

The presence of Newtonian gravity between bound objects implies that the geometry the internal observers will see is not flat but curved. The internal observers thus find a curved Lorentzian manifold. 

From the derivation above it is clear that we are dealing with a theory that goes beyond Newtonian gravity. To arrive at Newtonian gravity it was crucially important that we looked only at the quasi-static limit. Once the velocities of the bound objects are no longer small we have to take into account that the change of $\theta$ is not instantaneous. Gravity here has a finite propagation speed. 

Let us close this section with a remark on the equivalence principle. In equation (\ref{eqn:mass}) we have given a formula for the gravitational mass of a bound object. What remains to be shown is that the formula also gives the inertial mass of this bound object. The above derivation can be seen as a step in that direction. The excitations that bind together are all massless, move at the speed of light, and the notion of an inertial mass does not exist for them. To have an inertial mass we have to create a bound object from these excitations. But as we have just shown, a bound object implies a gravitational mass. To have an inertial mass thus implies the existence of a gravitational mass. What remains to be shown is that the two masses actually coincide. 

\section{Discussion}\label{sec:disc}
In this paper we proposed to change the relation that geometry and matter have with each other. The usual point of view is that geometry provides the stage on which matter propagates. In classical mechanics geometry provides an absolute frame. Whatever the matter does the geometry remains unchanged. The new element in general relativity is that geometry now reacts to the state of matter. Geometry tells matter how to move and matter tells geometry how to curve. The exact relation is given by Einstein's equations. Nonetheless,  geometry is still a stage for matter, albeit a dynamical one. 

Our objection to this setup is that one does not have direct access to the geometry; we  use matter to {\em infer} lengths and times. We believe it is desirable to have a theory where there is no geometry without matter, instead geometry and matter arise simultaneously. A theory where one does not act as the stage for the other. This is the kind of theory that we have proposed in this paper. Once we have matter, as emergent excitations of a more fundamental quantum system, this matter can be used to define geometrical notions. Using, e.g., the emergent gapless excitations as light rays one obtains the causal structure of the emergent theory. Geometry ceases to be a stage for matter and instead geometry and matter have a dual relationship to each other. One does not exist without the other. We call this theory \emph{Internal Relativity} and have seen that it naturally leads to a curved Lorentzian manifold. 

In section \ref{sec:gravity}, we relied heavily on the presence of an order parameter $\theta$ to derive Newton's law of gravitation. The argument for gravity seems to be more general though. If one has propagating excitations, then bound states of these excitations will attract each other. This can be be put more succinctly as follows:
\begin{center}
\emph{No propagation without gravitation.}
\end{center}
One might ask how it is that we can formulate a theory in which geometry and matter have this dual relationship. Why don't we need a background on which to formulate the theory? The key point here is that the relevant degrees of freedom of the fundamental theory are different from the matter degrees of freedom in the emergent theory. The same is true for the geometry. Geometric notions in the fundamental theory are different from the ones in the emergent theory. It is at this stage that emergence plays an important role. Because geometry and matter are emergent and not present in the fundamental theory they are able to influence each other.

In section \ref{sec:gravity}, we saw that a bound object has a gravitational mass given by 
\begin{equation}
m_g \simeq \int_{\partial C} (\nabla\theta)\cdot d\sigma.
\end{equation}
In section \ref{sec:ir}, we conjectured that the bound object has an inertial mass given by the same expression. This amounts to saying that the equivalence principle is a consequence of the theory and not an input. We have not shown that this is indeed the case but one can interpret the above result as pointing in the right direction. If one starts from massless excitations one does not have a notion of an inertial mass. As we have discussed in the end of section \ref{sec:gravity} to obtain such a notion one needs to make bound objects of these excitations which implies the existence of a gravitational mass. Thus the inertial mass and gravitational mass are closely linked. What remains to be shown is that they are given by the same expression.

The approach outlined here provides a new perspective on another problem in quantum gravity: the problem of time. If one performs a canonical analysis of pure general relativity one finds that the Hamiltonian vanishes. Instead of an evolution one finds a constraint. In a quantization of gravity one then faces the problem of having to reconstruct a spacetime picture from the timeless solutions to the Hamiltonian constraint. This problem is called the problem of time. The approach presented here shows this to be an unnecessary complication that arises because of an unphysical idealization that does not take into account that geometry and matter arise together. By neglecting one part, matter, and just focusing on the other part, geometry, one introduces the problem of time. The problem of time is the price one pays for not realizing that pure gravity is an unphysical idealization.

Another problem that is connected with viewing geometry as a stage for matter is the cosmological constant problem. If matter is viewed as propagating on geometry then the zero mode energy of matter fields should contribute to the curvature of geometry. This view leads to one of the worst predictions of theoretical physics. The cosmological constant obtained in this way is off by more then 120 orders of magnitude. The approach taken in this paper also sheds new light on this problem. In Internal Relativity matter is to be viewed as giving rise to geometry and so the above line of argument is seen to be fallacious. It is the excitations that make the geometry. Zero mode energies only appear in the effective description of the matter on a given spacetime. Fundamentally they should not be viewed as energies residing on the spacetime (see \cite{dreyerccp} for more details). 

Is the theory presented here an aether theory? The answer to this question is yes and no. Before special relativity the aether was introduced with one purpose: It was the carrier of electromagnetic radiation. In our theory the ground state characterized by the value $\theta_0$ plays a similar role. In this sense it is an aether theory. The old understanding of aether did not include other matter though. Matter was thought of as being different from the aether. People thought of matter as being immersed in a sea of aether. It is here that the old understanding ran into serious trouble. How could matter interact with electromagnetic radiation but not notice the presence of the aether? Our understanding differs from this old understanding in that all of matter is to be thought of as excitations of the aether. There is no matter distinct from the one that is carried by the aether. It is this crucial distinction that makes an aether acceptable. 

Wheelers dictum "geometry tells matter how to move and matter tells geometry how to curve" can be clearly observed in our theory. As the bound objects reacts to the gravitational force the distribution of $\theta$ changes. This in turn implies a change in the geometry. Back-reaction is not a stumbling block in our theory as it is in analog models of gravity \cite{analog}. 

Another point worth mentioning is that the speed with which gravity propagates is the same speed with which the other matter propagates. This is because both are the consequence of the same term in the Hamiltonian, namely
\begin{equation}
\int d^3x\;(\nabla\theta)^2.
\end{equation}
It is a satisfying feature of the model that it does not require any fine tuning to achieve this equality of propagation speeds. The finiteness of the propagation speed also shows that is more then just Newtonian theory. Only in the limit of small velocities does our theory reduce to the case of Newtonian gravity.  

In our current understanding of special relativity Lorentz symmetry is an a priori symmetry. Only fields constructed in such a way that they are Lorentz covariant are allowed in the theory. In this setup elementary particles are naturally viewed as the irreducible representations of the Poincar\'e group (see, e.g., \cite{weinberg}). In our approach the relationship between particles and symmetry group is exactly reversed. It is the particles that determine structures like the light cone and the symmetry group. We are thus proposing not to use the Poincar\'e group and its representation theory in the basic setup of the theory. 

Recently we proposed a new way to look at the measurement problem in quantum mechanics \cite{dreyerqm}. In this proposal classicality is a property of large quantum systems. The bound objects that we have been looking at in section \ref{sec:gravity} are of this type. Their rigidity is the property that makes them appear classical. Thus, it appears that there is  a connection between gravity and classicality. Classical behavior requires bound objects which in turn imply curvature. 

\begin{figure}
\begin{center}
\includegraphics[height=6cm]{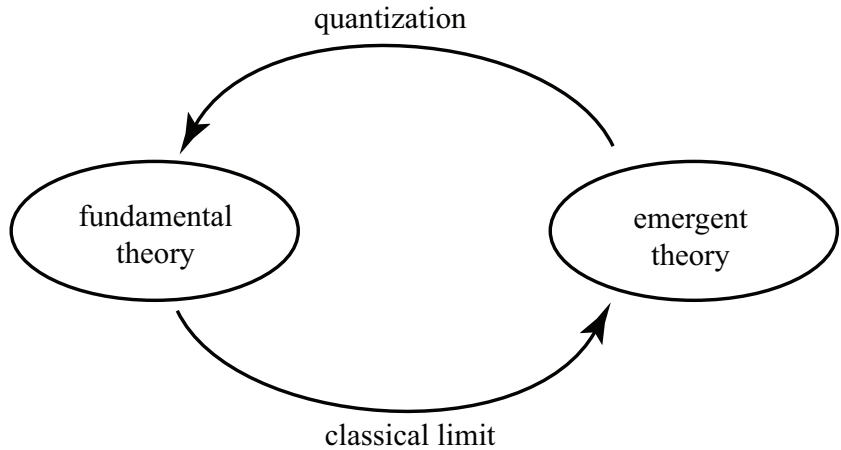}
\end{center}
\caption{When one quantizes a theory one assumes that the above circle closes, i.e. that the quantization of the classical limit of the fundamental theory again gives the fundamental theory. In the approach that we advocate here this is not the case. The fundamental theory is radically different from the quantization of the emergent theory.}\label{fig:quantize}
\end{figure}

It is important for our approach that geometry and matter are truly emergent. This implies that the fundamental theory can not be obtained through a process of quantization (see figure \ref{fig:quantize}). This is because quantization always results in a quantum theory in which the classical states survive as labels for quantum states. Applied to quantum gravity this implies that the fundamental theory should not be viewed as a theory of superimposed spacetimes. Instead the fundamental theory is free of geometric notions. They only arise later together with the matter degrees of freedom.

This should be viewed as an advantage of our proposal since it has been extremely difficult to construct Hilbert spaces of spacetime geometries or to make sense of superpositions of spacetimes (with the notable recent exception of CDT \cite{loll}). We also note that a new direction is developing in recent work, where  
traditional superpositions are being  abandoned in favor of a background structure (examples are the new approach of algebraic quantum gravity \cite {thomasaqg},  the computational universe \cite{seth} and quantum graphity \cite{KonMarSmo}). 

An interesting consequence of our point of view is that it allows for a new set of observable consequences. Recently the possibility of observing Lorentz violating effects as a consequence of quantum gravity has attracted a large number of researchers. Already available data together with experimental data now becoming available (some of which was presented in this conference) suggests that these effects might actually be too small to be observable. This is why our approach is interesting. It allows for two possible new areas of observable effects. One is related to the early universe the other to the size of natural constants. 

Since in our approach geometry is emergent, one can ask what the effects of emergent geometry are. What are the remnants of the process of the emergence of geometry? The hope would be that these remnants could be observed in the cosmological microwave background.

The fundamental constants of nature, like the gravitational constant $G$, Planck's constant $\hbar$, the speed of light $c$, and the fine structure constant $\alpha$, are emergent in our approach. They can all be expressed in terms of the parameters of the fundamental theory. In particular there ought to be relations between these constants that can not be understood in the emergent theory alone. One might, e.g., understand why gravity is so small compared to the other forces. Because the constants of nature appear to be fundamental if one has only access to the emergent theory the hierarchy problem is completely out of reach in the emergent theory. 

\begin{acknowledgments} This work has been supported by the European Science Foundation network programme  "Quantum Geometry and Quantum Gravity" and partially supported by a grant from the Foundational Questions Institute (fqxi.org). The author would also like to thank Lee Smolin, Fotini Markopoulou, and Chris Isham for fruitful discussions. Finally the author would like to thank the Imperial College London where this work was started and the Perimeter Institute for Theoretical Physics for hospitality during part of this research.
\end{acknowledgments}

\end{document}